\documentclass[graybox]{svmult}
\usepackage{mathptmx}
\usepackage{amsmath}
\usepackage{microtype}
\usepackage{float}
\usepackage{graphicx}
\usepackage{hyperref} 

\begin{document}
\title*{The Emission of Electromagnetic Radiation from Charges Accelerated by Gravitational Waves and its Astrophysical Implications}
\titlerunning{Radiation from Charges Accelerated by Gravitational Waves}
\author{{\bf Mitchell Revalski, Will Rhodes, and Thulsi Wickramasinghe}} 
\institute{Mitchell Revalski \and Will Rhodes \and Thulsi Wickramasinghe \at The College of New Jersey, 2000 Pennington Road, Ewing, New Jersey, 08628, USA \\ \email{revalsm1@tcnj.edu, rhodesw1@tcnj.edu, wick@tcnj.edu}}
\maketitle

\vspace {-25 mm} 
{\bf
This is part two of two. For part one see:\\
\href{http://adsabs.harvard.edu/abs/2015ASSP...40..295W}{http://adsabs.harvard.edu/abs/2015ASSP...40..295W}
}\\

\abstract{
We provide calculations and theoretical arguments supporting the emission of electromagnetic radiation from charged particles accelerated by gravitational waves (GWs). These waves have significant indirect evidence to support their existence, yet they interact weakly with ordinary matter. We show that the induced oscillations of charged particles interacting with a GW, which lead to the emission of electromagnetic radiation, will also result in wave attenuation. These ideas are supported by a small body of literature, as well as additional arguments for particle acceleration based on GW memory effects. We derive order of magnitude power calculations for various initial charge distributions accelerated by GWs. The resulting power emission is extremely small for all but very strong GWs interacting with large quantities of charge. If the results here are confirmed and supplemented, significant consequences such as attenuation of early universe GWs could result. Additionally, this effect could extend GW detection techniques into the electromagnetic regime. These explorations are worthy of study to determine the presence of such radiation, as it is extremely important to refine our theoretical framework in an era of active GW astrophysics.
}

\vspace {-5 mm} 

\section{Introduction}
\vspace {-2 mm} 
Gravitational waves (GWs) are ripples in space which propagate at the speed of light, conveying the changing gravitational field of a system. Systems undergoing anisotropic acceleration will have changing gravitational fields and emit GWs \cite{schutz00}. The theoretical framework for GWs has been well developed and clarified making the field of study more accessible \cite{thorne95}, with many astrophysical applications explored \cite{sath09}. Despite this theoretical competency, gravitational radiation is the last prediction of general relativity which has not yet been detected directly.\par

Significant indirect evidence has accumulated through monitoring orbital decay in binary star systems as predicted by general relativity \cite{weisberg10, hermes12}. This decay is attributed to orbital energy radiated in the form of GWs. Recently, {\it B}-mode polarization has been measured in the Cosmic Microwave Background (CMB), predicted if stochastic GWs in the early universe induced a curling polarization in the CMB \cite{ade14}. If confirmed, this will provide significant evidence for the existence of GWs.\par

The interaction of GWs with normal matter is extremely weak, with the graviton cross-section $\sim 80$ orders of magnitude smaller than the Thomson cross-section \cite{maggiore08}! Also, as compared to electromagnetic waves, the wavelengths of GWs are determined by the bulk movement of a source rather than atomic transitions. Thus GWs cannot be used to resolve sources in the way electromagnetic radiation is employed \cite{flanagan05}. Regardless, there is interest in GWs due to their pristine nature, propagating nearly unperturbed by galactic or interstellar media. Despite their weak interaction, it has been well established that GWs should cause a ring of particles to oscillate. This toy model is often the prototype for studying the interaction of particles with GWs \cite{hobson06}.\par

An interesting situation arises when considering a distribution of \textit{charged} test particles. If a GW causes a distribution of charged particles to oscillate, then they could experience an acceleration and radiate energy in the form of electromagnetic radiation. If this occurs, significant astrophysical consequences could result including attenuation of GWs and new avenues for GW detection.\par

We aim here to show that charges interacting with a GW will effectively accelerate, absorb GW energy, and radiate that energy in the form of electromagnetic radiation, leading to wave attenuation. The meaning of {\it effective} acceleration is detailed in these proceedings in the contribution by Thulsi Wickramasinghe where it is shown that a charged gas cloud under the influence of a passing gravitational wave displays the same time varying terms in the calculated magnetic potential as those produced by accelerating charges, leading to measurable radiation. Thus we are able to forgo an extensive investigation of whether or not charges experience the acceleration, as we find it leads to the same physics. We discuss the previous work on this topic in \S2, and present our calculations for the radiated power in \S3. We give additional arguments for acceleration in \S4, and devote \S5 to discussing the astrophysical implications. Our conclusions are given in \S6.

\vspace {-5 mm} 

\section{Previous Work}
\vspace {-2 mm} 
The concept of electromagnetic emission from charged particles accelerated by GWs is not new; however, it has been sparsely explored. Early work concluded that charged particles could extract energy from GWs and radiate it in the form of electromagnetic radiation \cite{heintz68}. Further work agrees, with the condition that a pre-existing magnetic field be present \cite{papa81}. Additionally, significant wave attenuation could occur under certain resonant conditions, allowing particles to extract large amounts of energy from the GW \cite{voyat06, kleidis93}. More recent work has examined these resonant and nonlinear interactions in magnetic fields, finding that charges are accelerated \cite{kleidis95}. In examining astrophysical situations in which these conditions may occur, it was shown that charges could be accelerated to high energies by the GWs emitted during the collapse of massive magnetized stars \cite{vlahos04}. Preliminary calculations for general distributions of particles show that while charged particles can absorb large amounts of energy from GWs, it would constitute at most 1 part in $\sim10^9$ of the wave's overall energy \cite{voyat06}.\par 

With this literature it is more firmly grounded that charged particles will, at least under some conditions, absorb energy from GWs, and emit electromagnetic radiation. The primary constraint in previous calculations is the requirement of a pre-existing magnetic field. We aim to show that charged particles will be accelerated by GWs and radiate in general without an external magnetic field, as further developed employing Maxwell's equations in the parallel paper led by Thulsi Wickramasinghe.\par

An important detail to carry through from this brief literature survey is that we expect charged particles will be accelerated by the GW with respect to their electrostatic fields, and not freely falling. Additionally, debates on if uniformly accelerating charges radiate do not apply here, as the direction of the acceleration is not constant for particles under the influence of a GW, meaning $\dddot {a(t)}$ will be nonzero.

\vspace {-5 mm} 

\section{Radiation from Accelerating Charges: Calculations}
\vspace {-2 mm} 
We turn now to calculating the power radiated by these charges, constructing an order of magnitude calculation from the non-relativistic Larmor formula. When the equations of motion for particles under the influence of a GW are examined in the transverse-traceless (TT) gauge, we may dispense with a relativistic treatment and work with Newtonian equations of motion as described in \cite{maggiore08, buon07}.\par

First, equations describing the acceleration for each particle in a ring distribution are derived. Following this we employ the Larmor formula to find a general expression for the total power radiated by all charges in a given distribution. We consider the total charge of the system to be constant in time. Using the Larmor formula, the power radiated from a single non-relativistic charge is
\begin{equation}
\label{larmor-eq}
P_e = \frac{1}{4\pi \epsilon_0}\frac{2}{3} \frac{e^2 a^2}{c^3}
\end{equation}
where $e$ is the particle's charge and $a$ is its acceleration. The acceleration can be calculated based on the particle's motion in the proper detector frame as derived in \cite{hobson06}. Considering motion in the plane perpendicular to the direction of the wave propagation vector, the particle's position for one wave polarization is given by
\begin{equation}
 X(t) = x\left(1+\frac{h}{2}sin(\omega t)\right)
\end{equation}
where $h$ and $\omega$ are the amplitude and angular frequency of the GW, respectively. $X(t)$ is the position of the particle in the proper detector frame and $x$ is its initial coordinate position. Differentiating twice with respect to time we obtain
\begin{equation}
\label{accel-eq}
\ddot{X}(t) = -\frac{1}{2} x h \omega^2 sin(\omega t)
\end{equation}
Substituting equation \ref{accel-eq} into the Larmor formula (\ref{larmor-eq}), we obtain the power emitted as a function of time.
\begin{equation}
P_e(t) =  \frac{1}{4\pi \epsilon_0}\frac{e^2}{6 c^3} x^2 h^2 \omega^4 sin^2(\omega t)
\end{equation}
The emitted power which is sinusoidal in time will have a continuous frequency distribution which may be obtained using Fourier techniques \cite{jackson98}. \par

To generalize our solution and make it astrophysically applicable, we extend this result to rings, disks, spheres, and cylinders of charges. Considering a ring in 2D space, we extend equation \ref{accel-eq} for both $\ddot{X}(t)$ and $\ddot{Y}(t)$ and obtain the proper acceleration, $\sqrt{\ddot{X}^2 + \ddot{Y}^2} = r\omega^2 h/2~ sin(\omega t)$, which is independent of the coordinate system. As the ring oscillates the charge density remains constant to first order in $h$, since the circumference of the ellipse, $4r(1+h/2~sin\omega t)E(2h~sin\omega t)$, is still $2 \pi r$. For a ring with total charge $Q = Ne$ consisting of $N$ point charges the total power radiated is
\begin{equation}
\label{power-ring-eq}
P_{ring}(t) = \frac{1}{4\pi \epsilon_0}\frac{2e}{3c^3} Q\ddot{r}^2 = \frac{1}{4\pi \epsilon_0}\frac{ e h^2  \omega^4 Q}{6c^3}  r^2 sin^2(\omega t)
\end{equation}
Expanding this to a disk of charges, we consider a series of concentric annuli with charge density $\sigma$, which is constant in time to first order in $h$. Integrating over these annuli yields a total charge on the disk of $Q_{disk} = (\pi r^2_2 - \pi r^2_1)\sigma$, where $r_1$ and $r_2$ are the inner and outer radii, respectively. The charge of each infinitesimally thin ring is $dQ = 2\pi r dr \sigma = 2Q_{disk} r dr/(r^2_2 - r^2_1)$. From equation \ref{power-ring-eq} we can write the power emitted by $dQ$ as
\begin{equation*}
dP = \frac{1}{4\pi \epsilon_0}\frac{e  h^2  \omega^4}{6c^3}  r^2 sin^2(\omega t)\; dQ
\end{equation*}
Substituting in $dQ$ and integrating over $r$, allowing $r_1 \rightarrow 0$ for a uniform disk, we find
\begin{equation}
\label{power-disk-eq}
P_{disk}(t) = \frac{1}{4\pi \epsilon_0}\frac{  e  h^2  \omega^4  Q_{disk}}{12c^3}r^2 sin^2(\omega t)
\end{equation}
where $r$ is the proper radius of the disk at $t=0$ and $Q_{disk}$ is the total charge. \par
Considering a spherical distribution, we sum our result from equation \ref{power-disk-eq} for different cross-sections of a sphere. At $t=0$, the proper radius of the sphere, $R$, is related to the cross-sectional radius by $R^2 = z^2 + r^2$. The total charge of the sphere is $Q_{sph} = 4\pi R^3\rho$/3, where the charge density $\rho$ is constant with time to first order in $h$. The charge for each infinitesimal cross-section is then $dQ = 4\pi r^2 \rho dz = 3Q_{sph} r^2 dz /4 R^3$. The power associated with this cross section from equation \ref{power-disk-eq} is
\begin{equation*}
dP = \frac{1}{4\pi \epsilon_0}\frac{  e  h^2  \omega^4}{12c^3} r^2sin^2(\omega t) dQ
\end{equation*}
Substituting in $dQ$, changing variables, and integrating from $z = -R \rightarrow R$, we find
\begin{equation}
P_{sph}(t) = \frac{1}{4\pi \epsilon_0} \frac{ e  h^2  \omega^4 Q_{sph}}{15 c^3}R^2 sin^2(\omega t)
\end{equation}\par
Far from a source, GWs may be treated as plane waves, propagating across a distribution of matter cylindrically. Thus we also consider a cylindrical distribution of charge with a base perpendicular to the GW propagation vector, $\vec z$. Using equation \ref{power-disk-eq} with amplitude $h(z) = h_1/z$, and integrating over the cylinder length, $z_1 \rightarrow z_2$, we obtain
\begin{equation}
\label{power-cyl-eq}
P_{cyl}(t) = \frac{1}{4\pi \epsilon_0}\frac{  e h_{1}^{2}  \omega^4  Q_{cyl}}{12c^3}r^2sin^2(\omega t)\;\frac{1}{z_1 z_2}
\end{equation}

Using the equations above, the power radiated from astrophysical systems is calculated. We consider a small spherical cloud of $10^{12}$ electrons; so that our approximations hold. Examples are given in the following table.
\renewcommand{\arraystretch}{1.4}
\setlength{\tabcolsep}{0.5em}
\begin{center}
  \begin{tabular}{| c | c | c | c || c | c | c | c |}
    \hline
    $h$ & $\omega$ & N & Power (J/s)    &    $h$ & $\omega$ & N & Power (J/s)\\ \noalign{\hrule height 2pt}
    $10^{-20}$    &    $10^{6}$    &    $10^{12}$    &    $\mathcal{O}(10^{-51})$   &        $10^{-5}$    &    $10^{6}$    &    $10^{12}$    &    $\mathcal{O}(10^{-21})$ \\ \hline
    $10^{-12}$    &    $10^{2}$    &    $10^{12}$    &    $\mathcal{O}(10^{-43})$   &        $10^{-12}$    &    $10^{10}$    &    $10^{12}$    &    $\mathcal{O}(10^{-27})$ \\ 
    \hline
  \end{tabular}
\end{center}
Integrating over the duration of the wave will increase the amount of energy absorbed and re-emitted during the process.  It is important to note that the GW wavelength limits the value of $r$, determining the maximum separation that two particles could have and still be treated with simplified Newtonian equations of motion.\par

For our first-order treatment here, we have considered the equations of motion to be linear, which holds for small amplitude GWs ($h<<1$). The amplitude is considered to be constant over the distance which the charge distribution occupies, which is reasonable far from the source. An exception is equation \ref{power-cyl-eq}, where we have done a more complete treatment. As mentioned earlier, the charge density is constant in time to first order in $h$. Additionally, to calculate the total power from any distribution with a significant length in the direction of wave propagation as compared with the wavelength of the GW, the phase of the wave plays an important role \cite{marsh11}. For simplicity, we neglect this minor effect which when averaged over only contributes a factor of $\sim 1/2$ to the power radiated.

\vspace {-5 mm} 

\section{Additional Arguments for Radiation: Memory Effects}
\vspace {-2 mm} 
To make a stronger case for charges experiencing acceleration, we briefly examine the various \textit{memory} effects of GWs. It has been established that all sources of gravitational radiation will produce some degree of memory effect \cite{favata10}.\par

The Velocity-Coded Memory effect is a permanent relative velocity between two bodies following the passage of a GW. This effect should present itself when there is an asymmetric rate in the rise and fall of the GW amplitude, as shown in figure one of \cite{grishchuk89}. This work showed the effect would be very small, but could be of experimental importance.\par

The Linear Memory effect permanently changes the separation of two bodies following the passage of a GW \cite{braginsky87}. If the separation changes permanently, the particles should have accelerated despite the initial and final velocities being zero.\par

The Nonlinear Memory component results in permanent displacements which may be only one order of magnitude weaker than the oscillatory portion of strong GWs \cite{christo91}. This memory effect surprisingly shows up at leading quadrupole order in post-Newtonian expansions \cite{favata10}, and is due to waves produced by the energy carried in the initial GWs, in the form of the emitted gravitons \cite{thorne92}.\par

While these memory effects could be used in detection, the resulting displacements are very small. The possible exception being velocity-coded memory, as the particles continue to increase in separation even after the wave has passed. These effects are likely undetectable in ground based interferometers as the test masses are not truly free falling, relying on damping systems to maintain a delicate equilibrium.\par

A final motion to consider is the acceleration of particles in the direction of wave propagation. While almost always neglected when studying particle interactions with GWs, as the displacement is exceedingly small, it is calculated to be nonzero \cite{braginskii85}.\par 

The results to exploit here are not the memory effects themselves, but the corresponding acceleration that particles interacting with GWs should experience due to these memory effects.

\vspace {-5 mm} 

\section{Astrophysical Implications}
\vspace {-2 mm} 
The orders of magnitude given in \S3 are very small, even when GWs with very large amplitudes and high frequencies are involved. For average parameters the energy radiated per second by the charge distribution is on the order of a CMB photon; however, peak values yield $\sim 10^{-10}~ J\cdot s^{-1}$. In terms of the overall power of the GW as derived from its flux in \cite{hobson06}, we find the average fractional power extracted from a disk of charges to be
\begin{equation}
	\boxed{\frac{P_{disk}}{P_{incident}} \approx 1.8 \times 10^{-70} \omega^2 Q_{disk}} 
\end{equation}
Currently, the likelihood of detecting such weak radiation is highly unlikely. \par

Current resources invested in GW detection focus primarily on instruments like the Laser Interferometer Gravitational Wave Observatory (LIGO). If emission of electromagnetic radiation is induced by GWs through the process explored above, it would open a new regime for GW detection. These new detection methods, along with the use of charged particle storage rings \cite{dong03} and atom interferometers \cite{dimop09}, could allow GWs to be exploited in new studies of the universe.\par

Furthermore, if charges are able to attenuate GWs more than is currently understood, significant consequences would result for early universe cosmology. GWs produced in the initial expansion of the universe were likely very powerful, and many free charges would have been present, possibly contributing to significant attenuation of these waves. Further cases of attenuation would result during supernovae explosions, for example. Calculations regarding the significance of this attenuation are a current focus in extending this work.

\vspace {-5 mm} 

\section{Conclusions}
\vspace {-2 mm} 
We have discussed how GWs should accelerate charged particles, leading to the emission of electromagnetic radiation and wave attenuation. Due to the complexities of analyzing particle motion in a general relativistic framework, there may be arguments to oppose this which we have not yet considered. Despite this, an examination using Maxwell's equations yield radiation from a gas cloud under the influence of a GW, even if the particles are not experiencing the tidal accelerations. By further exploring the memory effects of GWs, we lent additional evidence to the ability of GWs to accelerate charged particles.\par

Our preliminary calculations show that even under idealistic circumstances in which systems produce strong GWs, the emitted radiation would be feeble. This follows since if high quantities of radiation were being produced, the effect would have likely been detected by now.\par

The most important conclusion is the theoretical support that this effect may occur. While it likely contributes only in a minor way to wave attenuation, it is significant in forming a coherent and complete pedagogy for GW interactions in our universe.\par

In the future we hope to collaborate with others and challenge the ideas presented here in order to reach a more secure conclusion. During this exciting time in GW astronomy, explorations such as this are important in completing our understanding of GW astrophysics and for explaining future observational phenomena.

\vspace {-5 mm} 

\section*{Acknowledgements}
\vspace {-2 mm} 
We would like to thank the School of Science and the Department of Physics at The College of New Jersey for funding to present these results at the Sant Cugat Forum on Astrophysics held in Barcelona during 22-25 of April, 2014. In addition, we are very grateful to the forum organizers for partially subsidizing our registration.  This research has made use of NASA's Astrophysics Data System (ADS).

\vspace {-5 mm} 

\end{document}